\documentclass[sigconf,nonacm, screen=true]{acmart}

\AtBeginDocument{%
  }

\setcopyright{cc}
\copyrightyear{2024}
\acmYear{2024}
\setcopyright{rightsretained}
\acmConference[ICAIF '24]{5th ACM International Conference on AI in
Finance}{November 14--17, 2024}{Brooklyn, NY, USA}
\acmBooktitle{5th ACM International Conference on AI in Finance (ICAIF '24),
November 14--17, 2024, Brooklyn, NY, USA}
\acmDOI{10.1145/3677052.3698604}
\acmISBN{979-8-4007-1081-0/24/11}

\usepackage[ruled, linesnumbered, noend]{algorithm2e}
\SetCommentSty{small}
\DontPrintSemicolon
\usepackage{cleveref}
\usepackage[shortlabels]{enumitem}

\Crefname{figure}{Fig.}{Figs.}
\Crefname{algorithm}{Alg.}{Algs.}

\newcommand{\R}{\mathbb{R}}
\newcommand{\E}{\mathbb{E}}
\newcommand{\Var}{\mathbb{V}}
\newcommand{\Exp}[2][]{\mathbb{E}_{{#1}}\left[#2\right]}
\newcommand{\norm}[1]{\lVert #1 \rVert}
\newcommand{\eye}{\mathbb{I}}

\newcommand{\tr}{\text{tr}}

\newcommand{\tvec}[1]{\mathrm{vec}\left(#1\right)}
\newcommand{\tvectext}[1]{\mathrm{vec}(#1)}

\DeclareMathOperator{\diag}{diag}

\theoremstyle{plain}

\theoremstyle{definition}

\theoremstyle{remark}

\newcommand{\act}{u}
\newcommand{\actt}{u_t}

\newcommand{\der}{\mathcal{D}} 

\newcommand{\loss}{\mathcal{L}}

\newcommand{\jac}{{J}}
\newcommand{\curvmat}{B}
\newcommand{\curva}{\hat{B}}
\newcommand{\curvmatfamily}{\mathcal{B}}

\newcommand{\hess}{H}

\newcommand{\pf}{\Gamma}

\newcommand{\tc}{C}
\newcommand{\tcper}{c}

\newcommand{\info}{I}

\newcommand{\shrinkintensity}{\varrho}
\newcommand{\trustregion}{\rho^{\text{TR}}}

\begin{document}

\title{Fast Deep Hedging with Second-Order Optimization}

\author{Konrad Mueller}
\orcid{0009-0009-5856-7038}
\affiliation{%
  \institution{Imperial College London \\
  J.P. Morgan}
  \city{London}
  \country{UK}
}

\author{Amira Akkari}
\orcid{0009-0000-5951-2818}
\authornote{Equal contribution}
\affiliation{%
  \institution{J.P. Morgan}
  \city{London}
  \country{UK}
}

\author{Lukas Gonon}
\orcid{0000-0003-3367-2455}
\authornotemark[1]
\affiliation{%
  \institution{Imperial College London}
  \city{London}
  \country{UK}
}

\author{Ben Wood}
\orcid{0009-0007-3186-3094}
\authornotemark[1]
\affiliation{%
  \institution{J.P. Morgan}
  \city{London}
  \country{UK}
}

\renewcommand{\shortauthors}{Mueller et al.}

\begin{abstract}

Hedging exotic options in presence of market frictions
is an important risk management task. 
Deep hedging can solve 
such hedging problems
by training neural network policies
in realistic simulated markets.
Training these neural networks may be delicate and suffer from slow convergence, particularly for options
with long maturities and complex sensitivities
to market parameters.  
To address this,
we propose
a second-order optimization scheme
for deep hedging.
We leverage pathwise differentiability
to construct a
curvature matrix,
which we approximate
as block-diagonal and Kronecker-factored
to efficiently precondition gradients.
We evaluate our method on a challenging and practically important problem: hedging a cliquet option
on a stock with stochastic volatility by trading in the spot and vanilla options. 
We find that
our second-order scheme
can
optimize the policy in $1/4$ of the number of steps
that
standard adaptive moment-based optimization takes.
\end{abstract}



\maketitle

\section{Introduction}

Hedging financial derivatives is
a fundamental risk management problem.
Classical models provide solutions
by relying on strong simplifying assumptions
such as absence of market frictions.
Deep hedging \citep{buehler_2019_deep}
approximates solutions to
hedging problems
by training a neural network
to take optimal hedging decisions.
Deep hedging has gained popularity
in both academia and industry
as it applies to hedging problems
with an arbitrary contingent claim,
market simulator,
and hedging objective.
For instance,
deep hedging has been applied
to lookback options \citep{carbonneau_2021_deep}
and extended to
general initial portfolios
\citep{murray_2022_deep},
and
non-Markovian \citep{horvath_2021_deep}
or uncertain \citep{lutkebohmert_2022_robus}
simulators.
We refer to existing surveys \citep{hambly_2023_recen, ruf_2020_neura}
for a comprehensive overview.
In practice,
learning to hedge exotic options
is a challenging optimization
on a high-dimensional action space
that requires 
computing pathwise gradients
at high backpropagation depth.
Solving this optimization
reliably and quickly is practically important,
especially when refitting models
frequently to recalibrated
simulators.
Like most neural networks,
deep hedging models are
usually optimized with diagonally
preconditioned gradient descent
(e.g., Adam \citep{kingma_2015_adam}).
To improve optimization,
we propose a
\emph{second-order}
scheme that preconditions gradients
with the inverse of
a Kronecker-factored~\citep{martens_2015_optim}
curvature matrix.

The setting of our paper
is a challenging and practically important
hedging problem:
hedging a \emph{path-dependent} option
by trading in the spot and vanilla options.
While the focus of our paper lies on optimization,
we start with a modification to the
standard deep hedging framework
to efficiently parameterize option hedging.
We define a dynamically evolving action space
such that
puts and calls with
the same moneyness and maturity
are tradable on any path and at any time
(floating grid).
In this new formulation, representing
the current hedging portfolio is challenging,
but can be internalized into the neural network
through recurrent connections.

We then focus on
the deep hedging optimization.
Fundamentally, the optimization is difficult
because we can evaluate
the model-free
hedging error
only when the contingent claim matures.
This implies that
(i) to compute loss and gradients
we need to unroll the network over all time steps
and
(ii) credit assignment
to individual hedging decision
is difficult, especially since the
environment is highly stochastic.
To address these issues,
we propose a second-order optimization scheme
that builds on techniques
from both
reinforcement learning (RL)
and the training of
recurrent neural networks (RNNs).
Many popular RL algorithms
\citep{kakade_2001_a, schulman_2015_trust, wu_2017_scala}
are based on
natural gradient descent \citep{amari_1998_natur}
and
precondition policy gradients
with the inverse Fisher information matrix (FIM)
of the action distribution.
Preconditioning with the FIM
can be made scalable with
Kronecker-factored approximate curvature (KFAC)
\citep{martens_2015_optim},
which approximates the FIM
as block-diagonal
with Kronecker-factored blocks.
In contrast to most RL problems,
the deep hedging environment is differentiable,
so we do not need to randomize actions and
estimate gradients with the score function trick.
Consequently, preconditioning with the FIM is not applicable.

Instead, we construct a preconditioner
for the deep hedging problem
that leverages the
pathwise differentiable transition dynamics.
Our preconditioner respects the curvature
of the loss with respect to actions
but linearizes the neural network.
It is linked to the Hessian of the
deep hedging problem, as it approximates
a generalized Gauss Newton matrix \citep{schraudolph_2002_fast}.
We use the KFAC approximations
\citep{martens_2015_optim, martens_2018_krone, george_2018_fast}
on our preconditioner
to obtain a tractable algorithm.
To stably invert the curvature matrix,
we propose a
new shrinkage-based damping scheme.

We compare the resulting second-order scheme
with Adam
on training a neural network to hedge
a \emph{cliquet} option
in a stochastic volatility model.
With Kronecker-factored preconditioning,
the optimization progresses significantly
faster per-iteration.
This translates into a decrease in training time
as additional computational costs
can be amortized \citep{ba_2017_distr}.
We analyze the learned strategy
for hedging a cliquet and highlight
the importance of using options as hedging instruments.

\section{Deep Hedging}

\subsection{Hedging problem}

We seek to approximately solve
the problem of hedging
an exotic equity derivative
in finite, discrete time.
We consider
equidistant time steps
${t \in \{0, 1, \dots, T\}}$
up to a time horizon $T < \infty$,
and random variables
on a filtered probability space
${(\Omega, \mathcal{F}, \mathbb{F}, \mathbb{P})}$.
The filtration
$\mathbb{F} = (\mathcal{F}_t)_{t \geq 0}$
is generated as
${\mathcal{F}_t = \sigma(\info_0, \dots, \info_t)}$,
where
$\info_t$
captures newly available market information
at time $t$
and takes values in
$\mathbb{R}^{d_t}$.
We denote $\mathcal{F}_t$\nobreakdash-measurable
random variables with subscript~$t$.

Consider an underlying with
$\R_{> 0}$\nobreakdash-valued stock price process $(x_t)_{t \geq 0}$.
At time $t=0$ a bank sells
a path-dependent derivative for a price $p_0>0$. The derivative has payoff $\psi(x)$ at time $T$ for some 
$\psi: \R_{> 0}^{T + 1} \to \R$.
At times $t \in \{0, \dots, T - 1\}$,
the bank chooses an
$\R^d$\nobreakdash-valued control (also ``action'') $\actt$,
representing trades in
$d$ different hedging instruments.
The first component
$\actt^{(1)}$ is a trade
in the underlying $x$;
other hedging instruments
are European puts and calls on $x$.
We therefore focus on hedging spot and volatility risk,
and neglect other risk factors.
In particular, we assume deterministic interest rates 
and treat prices and payments
as discounted to $t = 0$.
We highlight two different modeling choices
for specifying which vanilla options
are available for trade at any time $t$
(let $i > 1$ below).
\begin{enumerate}[(a)]
    \item 
    \emph{Fixed grid:}
    $\actt^{(i)}$ is a trade
    in the same contract at each time~$t$.
    \begin{itemize}
        \item Example:
        If $\actt^{(i)}$ trades a call option with
        strike $K^i > 0$ and maturity $\tau_i$, then
        $\act_{t + 1}^{(i)}$ trades the same call option
        with strike $K^i$ and reduced
        time to maturity
        $\tau_i - 1$.
    \end{itemize}
    \item
    \emph{Floating grid:}
    $\actt^{(i)}$ is a trade in an option
    with the same moneyness and time to maturity at each time $t$.
    \begin{itemize}
        \item Example:
        For all $t$,
        $\actt^{(i)}$
        is a trade in a
        call option
        with time to maturity $\tau_i > 0$ and
        strike
        ${K^{i}_t = x_t \exp({k_i})}$,
        with moneyness 
        ${k_i \in \R}$.
    \end{itemize}
\end{enumerate}
\citet{buehler_2019_deep}
use a fixed grid
as they
consider trading in the same $d$ hedging instruments
at all times.
In this setting,
we can 
unwind past trades and
track the
hedging portfolio 
\begin{equation}\label{eq:hedging_portfolio}
    \vartheta_t = \sum_{s = 0}^t \act_s  ,
\end{equation}
as the sum
of past hedges.
The fixed grid
can accommodate all types of
hedging instruments
but assigns a new dimension
to each tradable contract.
To capture all relevant puts and calls
on any path $(x_t)_{t \geq 0}$ and at any time $t$,
we need to choose a large grid size $d$
that depends increasingly on $T$.

\begin{figure}
\centering
\includegraphics{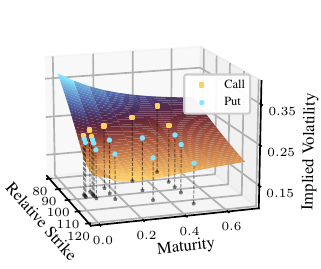}
\Description{
TODO: description 
}
\caption{
Floating grid.
Implied volatility surface is stochastic;
relative strike (in \%) and maturity
of tradable options are constant.
Marked puts/calls are used in
our experiments.
}
\label{fig:floating_grid}
\end{figure}
The floating grid addresses this issue 
because at each time step we only need to represent the much
smaller set of options that are relevant in the
current state. 
This floating definition of the action space means that 
each option can only be traded once, implying
that all trades are held
until maturity.
However, a trader can approximately offset an option traded at
time $t$ in later trading, as long as the option remains within 
the relevant set.\footnote{
For example consider buying a $40$\nobreakdash-day call at $t = 0$ with strike $K_0^i$.
After $20$ days we may offset the terminal payoff of this call option
by selling a convex combination of two $20$\nobreakdash-day calls
with $K_{20}^j \leq K_0^{i} \leq K_{20}^{l}$.
This requires the availability
of such calls.
}
The floating grid provides a reasonable representation of the 
typical range of available liquidity in practice.
We do not focus on the details of translating a floating grid policy
into real-world trades in this work.

We consider hedging on the floating grid
with the convention that
only out-of- or at-the\nobreakdash-money options are available.
If $k_i > 0$, the $i$th option is a call,
and
if $k_i \leq 0$, it is a put.
We remove options that mature after $T$ from the grid,
i.e., set $\actt^{(i)} = 0$ when $\tau_i > T - t$.

Trades $u_t$ increase or decrease the terminal PnL by
$\actt^\top r_{t, T}$,
where $r_{t, T}$ is the total return realized up to time $T$
of the $d$ instruments traded at time $t$.
The total return of trading the underlying $x$,
is given by ${r^{(1)}_{t, T} = x_T - x_t}$. For $i>1$ and
when the $i$th option is a call,
\begin{equation*}
    r^{(i)}_{t, T} = 
    (x_{t + \tau_i} - {K}^{i}_t)^+ - \mathcal{C}_t(\tau_i, k_i)
\end{equation*}
where $\mathcal{C}_t(\tau_i, k_i)$
is the premium paid for the
call at time $t$.
Put option returns are defined analogously
with payoff
$(K^i_t - x_{t + \tau_i})^+$.
All hedging leads to 
terminal gains (or losses) of
\begin{equation*}
   \pf_T(\act) := \sum_{t = 0}^{T - 1} \actt^\top r_{t, T},
\end{equation*}
and incurs transaction costs.
For simplicity,
we consider proportional transaction costs
that accumulate to
\begin{equation*}
    C_T(\act) :=
    \sum_{t = 0}^{T - 1} \tcper^\top \lvert \actt \rvert  ,
\end{equation*}
where costs $\tcper \in \R_{>0}^d$
are instrument-dependent,
and $\vert \cdot \vert$ is applied elementwise.
Hedges
are determined by a \emph{policy}
$\pi$
which is a collection of maps
$(\pi_t)_{t < T}$
such that
\begin{equation}\label{eq:policy_map}
   \actt = \pi_t \left (\info_0, \dots, \info_t \right),
\end{equation}
for all $t \in \{0, \dots, T - 1\}$.
We want to find a policy $\pi$
that minimizes
a given hedging objective,
for which there are plenty of well-motivated choices
\citep{buehler_2017_stati}.
In this work, we focus on minimizing
a weighted sum of PnL variance \citep{schweizer_1995_varia}
and transaction costs 
\begin{equation}\label{eq:deep_hedging_objective}
    \gamma \; \Var{
     \left(
     \pf_T(\act)
     - \psi(x) \right)
     }
     + \Exp{\tc_T(\act)}
     ,
\end{equation}
where actions $\act$ are given via the policy $\pi$
as in \cref{eq:policy_map}.
The risk aversion coefficient $\gamma > 0$
weighs the importance of
low hedging error vs. low transaction cost.

\paragraph{Cliquet}
Above, we described the problem of hedging
a generic European payoff $\psi$.
In our numerical experiments, we
take $\psi$ to be a 
\emph{cliquet option}.
Cliquets are traded in standard formats,
and are simple to describe,
but difficult to hedge because of their 
sensitivity to volatility dynamics.
This makes them a good test case for
deep hedging.
We consider a standard form of
the locally-capped, globally-floored
cliquet option, which pays at maturity~$T$
\begin{equation*}
\psi(x) = 
    \max \left[
    \sum_{\tau_i = \tau_1}^T 
    \min
    \left(
    \frac{x_{\tau_i}}{x_{\tau_{i - 1}}} - 1, \mu
    \right),
    0
    \right]
    ,
\end{equation*}
for reset dates $0 < \tau_1 < \dots < \tau_m = T$.
In each of the $m$ periods (typically of equal duration), we measure the return
of the underlying asset, apply a cap at $\mu > 0$; we then sum these returns, and 
floor the result at zero.
Intuitively, we can see that the contribution of the current period to the final 
payoff looks like a forward minus an option with moneyness 
(as of the period start) of $1+\mu$. However, we do not know the moneyness of 
the equivalent option in the next period, or the effect of the global floor.

\subsection{Parameterizing the hedging policy}\label{subsec:parameterizing_hedging}

Deep hedging \citep{buehler_2019_deep} approaches
the hedging problem by
parameterizing the policy with a neural network
and training it to minimize the hedging objective.
\citet{buehler_2019_deep} use the fixed grid formulation,
such that the path-dependence of optimal hedging decisions
(eq.~\ref{eq:policy_map})
reduces to 
a dependence on the portfolio $\vartheta_t$
when $I$ is a Markov process.
Then, they define the policy
through the parameterization\footnote{
On the fixed grid,
having the network output
$\actt$ or $\vartheta_t$ is
conceptually
equivalent.
}
\begin{equation}\label{eq:dh_policy}
    {\vartheta_t = \pi_\theta \left(\info_t, \vartheta_{t - 1} \right)},
\end{equation}
where $\pi_\theta$
is a neural network
with parameters $\theta \in \R^{n}$.
We slightly abuse notation here and
absorb time dependence into the
information process
(``$I_t \leftarrow (I_t, t)$'').
The network is recurrent in its output~$\vartheta_t$
but does not have a conventional hidden state.

For hedging policies on the floating grid,
this parameterization is not immediately applicable.
There,
trades $u_t$ and $u_s$ are in different contracts,
so that the hedging portfolio
cannot be computed as the sum of past hedges.
The precise portfolio at time $t$ can only be
represented as a concatenation of past hedges
${(\act_s)_{s \leq t}}$
and spot values
${(x_s)_{s \leq t}}$.\footnote{
The path of $x$ is needed
to track the strikes at which
past options were traded.
}
Such a naive representation is high-dimensional (growing in $T$),
sparse, and in our experience difficult to learn from.
Instead of tracking the portfolio precisely,
we represent it approximately as $h_t \in \R^k$
with $k \ll Td$.
Reducing the dimension of the portfolio representation
is less restrictive than reducing the dimension
of the action space
(e.g., using fixed grid with small $d$).
On any given path most actions will be close to zero,
such that the holdings' effective dimension is low.
This does not imply that these actions are useless;
option hedging makes a significant
contribution to the hedging performance
(see \cref{subsec:results}).

Option portfolios are commonly
represented via their 
Greeks~\citep{murray_2022_deep},
i.e., 
aggregated sensitivities of individual option prices
with respect to market parameters.
While low-dimensional, the Greeks representation
is expensive to compute in black-box simulators
and its compression may be limiting.
Instead, we learn the representation
through recurrent connections
and consider parameterizations
\begin{equation}\label{eq:floating_grid_map}
    \left(\actt, \{h^l_{t}\}_{l \leq L} \right)
    =
    \pi_\theta \left(I_t, \act_{t - 1}, \{h^l_{t - 1}\}_{l \leq L} \right)
    ,
\end{equation}
where $h^l_t$ is the hidden state of one of $L$
layers with a recurrent connection.
Unlike standard recurrent architectures
\citep{elman_1990_findi, hochreiter_1997_long},
this parameterization is additionally recurrent in
its outputs $(u_t)_{t < T}$.
This recurrence helps the network memorize
its output sequence:
the network can write $\act_{t - 1}$ to memory
during the time $t$ forward pass.
All recurrent connections
can also capture
non-Markovianity in
inputs $(I_t)_{t < T}$
\citep{horvath_2021_deep}.
Providing the last action as an input
is commonly done when training LSTMs
on RL problems.\footnote{
When using pathwise gradients,
gradients flow through
the action recurrence.
This is in contrast to training such networks
with policy gradient methods.
}
Removing the additional
recurrence in $u_t$
and treating deep hedging
as a sequence modeling problem,
is in our experience less efficient
but could allow for parallelization
with modern recurrent architectures~\citep{orvieto_2023_resur}.

\subsection{Optimizing the hedging policy}\label{subsec:optimizing_hedging_policy}

With a given parameterization,
the hedging problem reduces to a minimization
problem on the parameter space $\R^n$,
which can be solved
with gradient-based optimization.
The loss of \cref{eq:deep_hedging_objective}
is now a function
$\loss(\theta)$
of parameters,
whose gradient
can be estimated pathwise
\citep{mohamed_2020_monte}
as $\pi_\theta$ is differentiable in $\theta$
and the loss is pathwise differentiable in the
network outputs $(\actt)_{t < T}$.
Optimizing $\pi_\theta$ with
stochastic gradient estimates
and an optimizer such as Adam \citep{kingma_2015_adam}
works
but can be challenging.
A primary reason for this is that
computing pathwise gradients
requires backpropagating the error through time
and unrolling $\pi_\theta$ for $T$ times.
Even for small neural networks, this is
computationally expensive.
The common mitigation strategy of truncating
the backpropagation \citep{jaeger_2002_tutor}
is not applicable
as the hedging loss is
not \emph{separable} over time:
viewing the loss function
(eq. \ref{eq:deep_hedging_objective})
as a function of
$u$, we can write
the loss as
\begin{equation*}
    \sum_{t = 0}^{T - 1}
    \ell_t \left(\actt\right)
    +
    \ell_T
    \left(
    \sum_{t = 0}^{T - 1} \actt^\top r_{t, T}
    \right),
\end{equation*}
where the first term captures transaction costs.
The second term captures the quadratic hedging error
and is thereby non-linear in $u$
and can only be computed at time~$T$.
This is in contrast to
loss functions in supervised learning
of type ``$\sum_t \ell(u_t, y_t)$'',
where a component $\ell(u_t, y_t)$
can be evaluated immediately at time $t$.
This difficulty is
not specific to our loss function:
evaluating any measure of hedging error at $t < T$
requires valuing the portfolio,
which is model-dependent.
The loss decomposition also points to
an ill-conditioning issue:
$\actt$ impact the terminal loss component
only through the weighted sum $\pf_T$,
such that 
the \emph{per-sample} hedging loss
does not have a unique minimizer $u$.
While common in RL,
this is in contrast to
supervised learning problems,
where the ``square can usually be found inside
the sum'', i.e., $\ell(u_t, y_t) = (u_t - y_t)^2$.

Motivated by these observations, we
propose a second-order optimization scheme
for deep hedging.
Our second-order method is more efficient
on a per-step basis, addressing the
large per-step cost due to the
high backpropagation depth.
We precondition gradient descent updates
with a rich curvature matrix
that addresses the quadratic loss component $\ell_T$.
We replace the implicit loss linearization
of gradient descent
with a second-order model
that respects the curvature in~$\ell_T$
and only linearizes the neural network
$\pi_\theta$.

\section{Second-order Optimization for Deep Hedging}

\subsection{Second-order optimization of neural networks}

Before discussing second-order optimization for deep hedging,
we review the underlying concepts
in a standard supervised learning
setting
\citep{martens_2012_train, martens_2020_new}.
Given input $x$,
a neural network
$f_\theta$
outputs
$z = f_\theta(x)$
to match a target $y$
with
loss
$\loss(\theta) = \E[\ell(f_\theta(x), y)]$.
To minimize the loss,
we model
the loss landscape $\loss(\theta + \delta)$
for steps $\delta \in \R^n$
around the current iterate $\theta$ as
\begin{equation*} 
    M(\delta) =
    \loss(\theta)
    + \nabla \loss(\theta)^\top \delta
    + \frac{1}{2} \delta^\top \curvmat \delta ,
\end{equation*}
for some symmetric matrix $\curvmat \in \R^{n \times n}$.
If $\curvmat$ is positive definite, $M$ has a unique minimum at
$\delta^{*} = - \curvmat^{-1} \nabla \loss (\theta)$.
This is proportional to the step
chosen by
gradient descent,
Newton's method,
and
natural gradient descent \citep{amari_1998_natur},
when $\curvmat$
is the identity, Hessian,
or Fisher Information Matrix (FIM)
respectively.
In this work,
we focus on the class
of preconditioners $\curvmatfamily$
introduced by
\citet[sec. 12]{martens_2020_new},
where
any $\curvmat \in \curvmatfamily$
is of the form
\begin{equation*}
\curvmat =
\Exp[x, y]{
\jac_{z, \theta}^\top(x)
\hess(x, y, \theta)
\jac_{z, \theta}(x)
}
.
\end{equation*}
The distribution of $x, y$
can be arbitrary.
$J_{z, \theta}(x)$
is the Jacobian of the network
output with respect to $\theta$
evaluated at $z = f_\theta(x)$.
$H(x, y, \theta)$
takes values in
$\R^{m \times m}$, 
is positive definite,\footnote{
We assume this
for KFAC applicability;
the invariance only requires invertibility
\citep{martens_2020_new}.
}
and parameterization invariant
\citep[App. C]{kristiadi_2023_the}.
\citet[sec. 12]{martens_2020_new}
showed that preconditioned gradient descent
with any
$\curvmat \in \curvmatfamily$
satisfies an approximate invariance property.
The family
$\curvmatfamily$
generally does not contain the Hessian,
but it includes the FIM,
and the
\emph{extended} or
\emph{generalized Gauss-Newton}
matrix
(GGN)
\citep{schraudolph_2002_fast}.
The GGN
is a positive semi-definite (psd)
approximation of the Hessian matrix
and can be interpreted as the Hessian
of $\loss(\theta)$,
when replacing the inner model function $f_\theta$
with a ``linearization''
\citep{martens_2011_learn}.
Here, the GGN is the matrix
\begin{equation*}
G = 
\Exp[x, y]{
\jac_{z, \theta}^\top(x)
\hess_z(y)
\jac_{z, \theta}(x)
}
,
\end{equation*}
where $H_z(y)$ is the Hessian of
$\ell(z, y)$ with respect to model outputs~$z$.
The GGN is psd as $\ell$ is assumed to be
convex in $z$.

Practically,
preconditioners $\curvmatfamily$
are relevant because
we can approximate them efficiently
by imposing a Kronecker-product structure
(see \cref{subsec:kfac}).
This Kronecker-factored
approximation was first developed for the FIM
\citep{martens_2015_optim}
and exploits that the FIM
is the uncentered covariance matrix of gradients.
More generally, we can interpret \emph{any}
$\curvmat \in \curvmatfamily$
as a covariance matrix;
not of the actual gradient, but 
of a random vector
called the \emph{pseudo-gradient}
\citep{grosse_2022_chapt}.
Specifically, since
$H(x, y, \theta)$ is positive definite for any
$(x, y, \theta)$,
it is the covariance matrix of the random vector
$L Z$,
where
$L L^\top = H(x, y, \theta)$
and $Z$ has zero mean and identity covariance
(e.g., standard Gaussian).
Any
$\curvmat \in \curvmatfamily$ can be written
as the covariance matrix
\begin{equation*}
\curvmat =
\E_{x, y, Z}
\biggl [
\underbrace{
\vphantom{
\left(
\jac_{z, \theta}^\top(x)
L(x, y, \theta)
Z
\right)^\top
}
\jac_{z, \theta}^\top(x)
L(x, y, \theta)
Z
}_{\der \theta}
\;
\underbrace{
\left(
\jac_{z, \theta}^\top(x)
L(x, y, \theta)
Z
\right)^\top}_{
(\der \theta)^\top
}
\biggr ]
,
\end{equation*}
of the
random vector $\der \theta$.
We can sample $\der \theta$
by sampling $x, y, Z$,
computing $z = f_\theta(x)$,
$L$,
and \emph{backpropagating}
${\ell^{\text{pseudo}} = \langle z, L Z\rangle}$.
Consequently,
we can define the pseudo-gradient
$\der Y \in \R^{M \times N}$
for any matrix
$Y \in \R^{M \times N}$
as
${(\der Y)_{i, j} =
\partial \ell^{\text{pseudo}}} / \partial{Y_{i, j}} $.

\subsection{Kronecker-factored approximations}\label{subsec:kfac}

Computing, storing, and inverting
a dense preconditioner
${B \in \R^{n \times n}}$
is infeasible in deep learning.
But even simple, especially diagonal,
approximations of $\curvmat$
are useful:
Adam
\citep{kingma_2015_adam}
is motivated as a
``conservative'' (due to the square root)
approximation of the
empirical FIM's diagonal.
For \emph{feed-forward} neural networks,
a more accurate yet computationally feasible
approximation of any preconditioner
$\curvmat \in \curvmatfamily$
can be obtained through
Kronecker-factored Approximate Curvature (KFAC)
\citep{martens_2015_optim}.
KFAC provides a tractable preconditioner
$\curva \approx \curvmat$
through two major approximations:
\begin{enumerate}[(a)]
    \item $\curva$ is \emph{block-diagonal},
    where each block corresponds to weights
    of a single layer, allowing
    for a blockwise inversion of $\curva$.
    \item Each block $\curva_l$ is the
    \emph{Kronecker-product} of two smaller matrices,
    allowing for a
    factor-wise inversion of
    $\curva_l$.
    \label{enum:item:kfac}
\end{enumerate}
For \ref{enum:item:kfac},
consider the $l$th layer of $f_\theta$ with weight matrix
${W \in \R^{n_l \times n_{l-1}}}$.
Given an input $a \in \R^{n_{l-1}}$,
the layer computes pre-activation
${s = W a}$
and outputs $\sigma(s)$,
where $\sigma$ is the activation function.
The pseudo-gradient $\der W$ is given by
${\der W = g a^\top}$, where
${g = \der s}$
is the pseudo-gradient with respect
to $s$.
KFAC approximates the block as
\begin{align*}
    B_{l}
    &= \Exp{\tvec{\der W} \tvec{\der W}^\top}
    = \Exp{
    (a a^\top) \otimes
    (g g^\top)
    }\\
    &\approx
    \Exp{
    a a^\top
    }
    \otimes
    \Exp{
    g g^\top
    }
    = A \otimes G
    ,
\end{align*}
which can be viewed as an independence assumption
between $a$ and $g$,
with uncentered covariance matrices
$A := \mathbb{E}[aa^\top]$
and
$G := \mathbb{E}[gg^\top]$.
With this approximation,
any vector $\tvectext{V}$
can be efficiently preconditioned
by $B_l^{-1}$
via the identity
\begin{equation}\label{eq:kfac_multiply}
B_l^{-1} \tvectext{V}
= \tvectext{G^{-1} \; V \; A^{-1}}
.
\end{equation}
We estimate
$A$ and $G$
as exponential moving averages
and periodically compute updates to $A$
on an entire batch
and updates to $G$ on a single sample
(see lines
\ref{alg:line:a_update} and
\ref{alg:line:g_update}
in \cref{alg:kfac}).

\paragraph{RNN-KFAC \citep{martens_2018_krone}}
The Kronecker-product approximation
follows from $\der W = g a^\top$,
which holds when $W$ is used once during the
forward pass.
In recurrent architectures,
any weight matrix $W$ is applied repeatedly
to compute a new pre-activation
${s_t = W a_t}$
from input $a_t$
at any
${t \in \{0, \dots, T - 1\}}$.
The pseudo-gradient
\begin{equation*}
   \der W
   = \sum_{t = 0}^{T} g_t a_t^\top ,
\end{equation*}
is then a sum 
of gradient contributions 
${w_t := \tvec{g_t a_t^\top}}$, such that
\begin{equation*}
    \curvmat_l
    = \Exp{
    \left( \sum_{t = 0}^{T} w_t \right)
    \left( \sum_{t = 0}^{T} w_t \right)^\top
    }
    = \sum_{t = 0}^{T}
    \sum_{s = 0}^{T}
    \Exp{w_t w_s^\top}
    .
\end{equation*}
The analogous KFAC approximation
\begin{equation*}
    \Exp{w_t w_s^\top}
    = \Exp{(a_t a_s^\top) \otimes (g_t g_s^\top)}
    \approx A_{t, s} \otimes G_{t, s} ,
\end{equation*}
with
$A_{t, s} := \Exp{a_t a_s^\top}$
and $G_{t, s} := \Exp{g_t g_s^\top}$,
results in the approximation
\begin{equation}\label{eq:fisher_block_rnn_kfac}
\curvmat_l
\approx
\sum_{t = 0}^{T} \sum_{s = 0}^{T}
A_{t, s} \otimes G_{t, s}
,
\end{equation}
for the preconditioner block.
This approximation is a sum of Kronecker-products,
which we cannot invert efficiently.
To obtain a tractable approximation of $\curvmat_l$,
we follow \citet{martens_2018_krone} and assume
\begin{enumerate}[(a)]
    \item intertemporal independence:
    $A_{t, s} \otimes G_{t, s} = 0$, when $t \neq s$.
    \item stationarity:
    $A_{t, t} = A_{s, s}$
    and
    $G_{t, t} = G_{s, s}$
    for all $t, s$.
    \label{enum:item:stationarity}
\end{enumerate}
This yields the single Kronecker product approximation
\begin{equation}\label{eq:rnn_kfac_approx}
    \curvmat_l \approx T \; A_{0, 0} \otimes G_{0, 0}.
\end{equation}
Following \ref{enum:item:stationarity},
we approximate $A_{0, 0}$ with
samples of $A_{t, t}$ from all~$t$
(same for $G_{0, 0}$).
Even with these crude approximations,
\citet{martens_2018_krone}
show that the resulting
preconditioner performs well.
Below, we drop the indices and refer to
$A_{0, 0}$ as $A$ and $G_{0, 0}$ as $G$.

\paragraph{EKFAC \citep{george_2018_fast}}
Computing a dense
preconditioner $\curvmat$
is impractical.
However,
the diagonal of $B$ or $QBQ^\top$,
where $Q$ is an orthogonal matrix,
can be estimated efficiently
as each entry
is the variance of a (rotated)
partial pseudo-derivative.
Based on this observation,
\citet{george_2018_fast} 
improve the KFAC approximation
by re-estimating the diagonal
of $\curva_l$ in its eigenbasis.
To start, consider the eigendecomposition
of $\curva_l$
\begin{equation*}
    A \otimes G =
    Q \Lambda Q^\top = 
    (Q_A \otimes Q_G)
    (\Lambda_A \otimes \Lambda_G)
    (Q_A \otimes Q_G)^\top ,
\end{equation*}
where
$A = Q_A \Lambda_A Q_A^\top$,
and
$G = Q_G \Lambda_G Q_G^\top$.
We see that the Kronecker-product structure transfers
to $Q$ and $\Lambda$.
Because the diagonal of $\curvmat_l$ can be estimated cheaply, 
we can
remove the Kronecker-product assumption on $\Lambda$
and fit a general diagonal matrix $\tilde{D}$
\begin{equation}\label{eq:ekfac_min}
   \min_{\tilde{D} = \diag(\tilde{d}_1, \dots, \tilde{d}_{n_l \times n_{l - 1}})}
   \; \norm{
   \curvmat_l - Q \tilde{D} Q^\top
   }_F^2 ,
\end{equation}
to best approximate $B_l$ in the Frobenius norm.
The minimization
is solved by
$d_i^* = q_i^\top B_l q_i$,
where  $q_i$ denotes the $i$th column vector of $Q$
\citep{george_2018_fast, ledoit_2021_shrin}.\footnote{
$d_i^*$
not only minimizes the Frobenius loss,
but several sensible loss functions \citep{ledoit_2021_shrin}.
}
This ``solution'' is inaccessible
as $B_l$ is unknown
but $d_i^*$
can be replaced by a Monte Carlo approximation $d_i$
during the backward pass of
pseudo-gradients
(see eq. \ref{eq:ekfac_scale}).
The EKFAC approximation can be summarized as
\begin{align}\label{eq:ekfac_scale}
    \curva_l &= Q
    \diag\left({{d}_1, \dots, {d}_{n_l \times n_{l - 1}}}\right)
    Q^\top ,
    \nonumber
    \\
    \text{where}
    \quad Q &= Q_A \otimes Q_G ,
    \nonumber
    \\
    d_i &\approx d^*_i = \Exp{
    \left(q_i^\top \tvec{\der W}\right)^2
    } .
\end{align}
EKFAC was first introduced
in the context of correcting the
KFAC approximation of the FIM
of a probability distribution, parameterized
with a feed-forward or convolutional neural network.
We highlight that
EKFAC is applicable to
(i)~any $\curvmat \in \curvmatfamily$
and
(ii)~neural network architectures
with weight-sharing.
Consequently,
EKFAC also corrects
for the additional RNN\nobreakdash-KFAC assumptions
(eq. \ref{eq:rnn_kfac_approx}).

\paragraph{Damping scheme}

\begin{figure}
\centering
\includegraphics{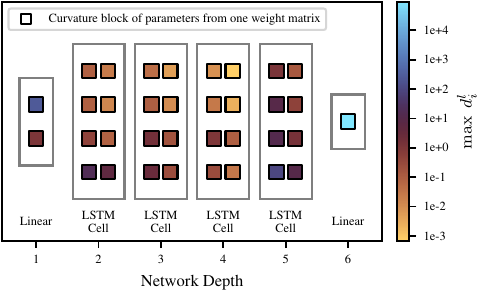}
\Description{
TODO: description 
}
\caption{
Magnitude of largest eigenvalues
differs across blocks
(end of training).
}
\label{fig:archtecture_max_scales}
\end{figure}

We can invert the
EKFAC approximation
via its eigendecomposition:
$\curva_l^{-1} = Q{D}^{-1}Q^\top$.
Inverting $\curva$
directly is usually unstable,
as even the true preconditioner often has many eigenvalues
close to zero.
It is common practice, to
instead invert
$\curva + \lambda \eye$,
where the damping hyperparameter $\lambda > 0$ often has
a significant effect on optimization performance
\citep{martens_2012_train, martens_2015_optim, ba_2017_distr}.
We find that on our problem and architecture,
choosing the same coefficient $\lambda$ for each block
$\curva_l$
leads to
over- or under-regularization
of different blocks,
as the
scale of the blocks' eigenspectra varies
significantly among them
(\cref{fig:archtecture_max_scales}).
Further, we do not find the adaptive
damping scheme of
\citet[App. C]{george_2018_fast} 
to alleviate the issue sufficiently.
The scheme uses
true eigenvalues
$\Lambda_A, \Lambda_G$
to adjust the effective damping strength.
We propose a new damping scheme,
that is based on the optimal scale ${D}$ instead.

As any $\curvmat \in \curvmatfamily$
is a covariance matrix,
we view damping
as 
\emph{shrinking} a covariance estimator
\citep{ledoit_2004_a}.
For a block-diagonal estimator $\curva$,
we suggest shrinking any block
$\curva_l \in \R^{N \times N}$
linearly to
\begin{equation}\label{eq:shrinkage}
    \curva_l \leftarrow (1 - \shrinkintensity) \curva_l +
    \shrinkintensity \frac{\tr(\curvmat_l)}{N}  \eye_N .
\end{equation}
The shrinkage target
$\tr(\curvmat_l) / N$
is block-dependent,
while
the shrinkage intensity $\shrinkintensity \in (0, 1)$
is a shared hyperparameter.
Because
\begin{equation*}
    \tr(\curvmat_l) = \tr(Q^\top \curvmat_l Q)
    = \tr({D^*})
    \approx \tr({D}),
\end{equation*}
we can estimate the trace
without imposing the Kronecker-product assumption.

\subsection{Preconditioning the deep hedging problem}

\begin{algorithm}
\caption{KFAC for Deep Hedging (DH-KFAC)}\label{alg:kfac}
\SetKwInput{Input}{Input}
\SetKwInput{Initialize}{Initialize}
\Input{samples $(I_t)_{t \geq 0}$;
$N^{\text{cov}}$, $N^{\text{evd}},
\beta^{\text{TR}},
\trustregion_0,
\beta^F,
\beta^D,
\beta^\text{mom}
$
}
\Initialize{
$\theta$, $\nabla_l^{\text{avg}}$, $A_l$, $G_l$, $D_l$ \footnotemark, $\trustregion$
}
\For{each iteration $i$ and mini-batch of paths $(I_t)_{t \geq 0}$}{
Compute $\act = (\actt)_{t < T}$ with $\pi_\theta$ on each $(I_t)_{t<T}$ \;
\If(\tcp*[f]{during forward}){$i \bmod N^{\text{cov}} = 0$}{\For(\tcp*[f]{update $A_l$}) {each layer $l$}{
$A_l \leftarrow \beta^F A_l + (1 - \beta^F) T^{-1/2} \sum_t \E_{\text{batch}}[a_t^l {a_t^l}^\top]$ \label{alg:line:a_update} \;
        }
    }
Sample a \emph{single} path with actions $u$ from mini-batch \;
    \textcolor{ACMDarkBlue}{Compute $H_u$ and Cholesky factor $L$} \tcp*{\cref{eq:action_hessian}} \label{alg:line:action_hessian}
    $\ell^{\text{pseudo}} \leftarrow \langle L Z, \act \rangle$, where $Z \sim N(0, \eye)$\;
    $g \leftarrow \nabla_\theta \ell^{\text{pseudo}}$\;
    \For(\tcp*[f]{update $G_l$ and $D_l$}){each layer $l$}{
    $G_l \leftarrow \beta^F G_l + (1 - \beta^F) T^{-1/2} \sum_t g_t^l {g_t^l}^\top$ \label{alg:line:g_update}\;
    $D_l \leftarrow \beta^D D_l + (1 - \beta^D) \left( Q_{G_l}^\top \text{mat}(g_l) Q_{A_l} \right)^2$\;
        }
\If{$i \bmod N^{\text{evd}} = 0$}{
    Compute eigenvectors $Q_{A_l}$ and $Q_{G_l}$ from $A_l$ and $G_l$\;
}
    
    Compute loss $\loss$ and gradient $\nabla \leftarrow \nabla \loss(\theta)$\tcp*[f]{backward}
    
    \For(\tcp*[f]{precondition gradient}){each layer $l$}{
        $\nabla^{\text{pre}}_{l} \leftarrow Q_{G_l}^\top \text{mat}(\nabla_{l}) Q_{A_l}$\;
        $\nabla^{\text{pre}}_{l} \leftarrow \nabla^{\text{pre}}_l \oslash \textcolor{ACMDarkBlue}{\left( (1 - \shrinkintensity) D_l + \shrinkintensity \text{ mean}(D_l) \right)}$ \label{alg:line:damping} \;
        $\nabla^{\text{pre}}_{l} \leftarrow Q_{G_l} \nabla^{\text{pre}}_l Q_{A_l}^\top$\;
    }
    
    $\eta \leftarrow \min \{ \sqrt{ \trustregion / \sum_l \langle \text{vec}( \nabla^{\text{pre}}_{l} ), \nabla_{l} \rangle }, \eta^\text{max} \}$ \label{alg:line:step_size} \;
    $\trustregion \leftarrow \beta^{\text{TR}} \trustregion$ \label{alg:line:tr_resizing}\tcp*{decrease trust-region \citep{ba_2017_distr}}
    $\nabla_l^{\text{avg}} \leftarrow \beta^{\text{mom}} \nabla_l^{\text{avg}} + \nabla_l^{\text{pre}}$\tcp*{momentum}
    $\theta \leftarrow \theta - \eta \nabla^{\text{avg}}$\tcp*{update parameters}
}
\end{algorithm}
\footnotetext{Here, $D_l$ is a dense matrix of shape $n_l \times n_{l - 1}$, as storing this is more efficient than a diagonal matrix of shape $(n_l n_{l - 1}) \times (n_l n_{l - 1})$ as in Equation~(\ref{eq:ekfac_scale}).}

We now construct
a second-order optimization scheme
for deep hedging
by preconditioning with
a Kronecker-factored matrix.
Unlike the related RL setting
\citep{kakade_2001_a, schulman_2015_trust, wu_2017_scala},
preconditioning with
the FIM is not applicable here,
as actions are not randomized,
i.e., $\theta$ does not parameterize
a probability distribution.\footnote{
In deep hedging, the optimal policy is deterministic, and 
the availability of analytical loss function gradients removes the 
training advantages of probabilistic actions.
}
However, we can construct
a GGN preconditioner
by decomposing the computation of $\loss(\theta)$
into a
model- and a loss function.
A valid GGN decomposition can be defined by treating
the tuple $\{\pf_T$, $C_T\}$
as the model output
because the per-sample
loss function is then convex in this model output.
In practice,
we find that this decomposition leads to subpar results
as the inner Hessian
is of rank 1.
To obtain a richer curvature model,
we define the
$d \times T$-~dimensional
vector of concatenated hedges
${\act = \{\act_0, \dots, \act_{T - 1}\}}$
as the model output.
When viewing
\cref{eq:deep_hedging_objective} as a function of
$\act$ directly,
the per-sample loss function is convex in $\act$.
Note that
this neglects that
through the action recurrence,
components in the output vector $\act$
\emph{depend} non-linearly on each other.
This additional approximation
is in the same spirit as
standard GGN approximations
\citep{schraudolph_2002_fast, martens_2011_learn}
as we ignore all second-order interactions that arise
within the network.

For the gradient, we use
an L1 transaction costs model
(eq. \ref{eq:deep_hedging_objective})
to reflect real-world market frictions.
For the inner Hessian,
a differentiable
L2\nobreakdash-type transaction costs model
is more suitable (better conditioning).
We therefore compute 
the inner Hessian not from \cref{eq:deep_hedging_objective},
but from the pathwise surrogate loss
\begin{equation*}
\ell(\act) = 
\gamma
\left(
\pf_T(\act)
- 
\psi(x)
- M
\right)^2
+
\sum_{t = 0}^{T - 1}
\sum_{i = 1}^d \tilde{c}^{(i)}
(\actt^{(i)})^2,
\end{equation*}
where $M$ is the mini-batch average of the hedged PnL
$(\pf_T(\act) - \psi(x))$,
whose dependence on the single path with action $\act$
we neglect.
The vector $\tilde{c} \in \R^{d}$
denotes the instrument-wise
L2 transaction costs,
which we set to $\tilde{c} = 8 c$ in our experiments.
We denote by $H_{\act}$
the Hessian of $\ell(\act)$ and by
$B^{\text{DH}}$
the proposed GGN\nobreakdash-type preconditioner given by
\begin{align} 
    B^{\text{DH}}
    &=
    \Exp{
    \jac_{\act, \theta}^\top
    H_{\act}
    \jac_{\act, \theta}
    }
    \nonumber
    \\
    H_{\act}
    &= 2 \gamma
    \;
    r_{\cdot, T}
    (r_{\cdot, T})^\top
    +
    2
    \diag{\left(
    [\tilde{c}, \dots, \tilde{c}]
    \right)}
    \label{eq:action_hessian}
,  
\end{align}
where $J_{a, \theta}$ is the
(pathwise)
Jacobian of the output vector $\act$
with respect to $\theta$,
and
$r_{\cdot, T} = \{r_{0, T}, \dots, r_{T - 1, T}\}$
denotes the concatenated returns corresponding
to the concatenated actions $\act$.

Our proposed preconditioner is
not the only reasonable choice
from $\curvmatfamily$,
but it outperforms
alternatives we tested in
preliminary experiments.
Alg.~\ref{alg:kfac}
summarizes
our second-order optimization scheme for deep hedging
(DH\nobreakdash-KFAC). 
Most individual steps are well known from KFAC
\citep{martens_2015_optim, ba_2017_distr},
EKFAC \citep{george_2018_fast},
RNN\nobreakdash-KFAC \citep{martens_2018_krone},
or pseudo-gradient sampling \citep{grosse_2022_chapt},
but their composition is new.
Other novel contributions include
the deep hedging
preconditioner used in line~\ref{alg:line:action_hessian}
and the damping scheme in line~\ref{alg:line:damping}.
The step size selection
with decaying trust-region size
(lines~\ref{alg:line:step_size}-\hspace{1sp}\ref{alg:line:tr_resizing})
is from \citet{ba_2017_distr}.

\section{Experiments}\label{sec:experiments}

\subsection{Setup}

\begin{figure}
\centering
\includegraphics{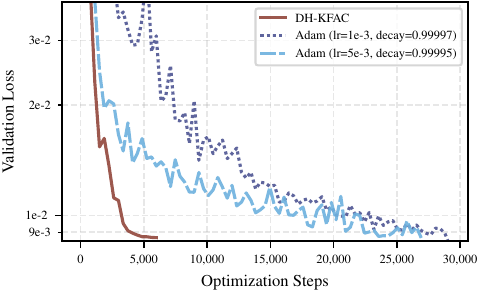}
\Description{
TODO: description 
}
\caption{
Validation loss during training.
}
\label{fig:loss}
\end{figure}

In this section, we
compare our second-order scheme
with Adam~\citep{kingma_2015_adam}
on training a deep hedging model.
We consider hedging a cliquet option in the
Heston model \citep{heston_1993_a}
and analyze the resulting hedging strategies.
With stochastic volatility
we can test how well the model learns to
use options to hedge volatility risk. 
In our discrete time setting
with transaction costs,
perfect hedging is not possible.
In the Heston model the price process and volatility satisfy 
\begin{align*}
    dx_s &= x_t \sqrt{v_t} dB_s \\
    dv_s &= \kappa(\theta - v_s) ds + \xi \sqrt{v_s} dW_s,
\end{align*}
where we assume no drift
and choose parameters
$x_0 = 1$,
$\kappa = 8$,
$v_0 = \theta = 0.0625$,
$\xi = 1.0$,
and correlation
$\rho = - 0.7$
for
Brownian motions $B$ and $W$.
For training, we simulate
$700{,}000$ paths
\citep{broadie_2006_exact}
for $T = 240$ steps, where
each step corresponds to a time
increment of $\Delta s = 1 / 250$.
At each step, we price
the $19$ vanilla options on the
floating grid (\cref{fig:floating_grid})
using the \texttt{StochVolModels}
library \citep{sepp_2024_logno}.\footnote{
In the format
$\{ \tau_i: K^i / x_t \}$
with multiple strikes per maturity,
the grid is:
$\{10: [0.99, 1.0, 1.01]\},
\{20: [0.97, 0.99, 1.0, 1.01, 1.03]\},
\{40: [0.95, 1.0, 1.05]\},
\{80: [0.91, 1.0, 1.09]\},
\{120: [0.85, 0.95, 1.0, 1.05, 1.15]\}$.
}
The cliquet has maturity~$T$,
cap ${\mu = 0.015}$,
and reset dates
$\tau_1 = 20, \tau_2 = 40, \dots, \tau_{12} = T$.
We assume transaction costs of
$c^{(1)} = 1\text{e-}4$,
and
$c^{(i)} = 1\text{e-}2$
for $i > 1$,
reflecting that delta hedging is cheaper than
hedging in options.
We set the risk aversion
to $\gamma = 1000$.\footnote{
The choice of the risk aversion coefficient $\gamma$
determines what hedging strategies are optimal
and it has to be tuned to underlying preferences in practice.
We chose $\gamma$ such that the learned hedge
clearly shows the impact of option hedging:
hedging policies trained with a smaller $\gamma$
rely more on cheap delta hedging.
With our choices for $\gamma$ and $c$, transaction costs
make up roughly $20 \%$ of the loss at the end of training.
}

As outlined in \cref{subsec:parameterizing_hedging},
we train a recurrent neural network
with additional action recurrence
(eq. \ref{eq:floating_grid_map}).
Input features $I_t$ consist of
an encoding of time $t$,
spot values
$x_t$ and $x_{\max \{\tau_i \leq t\}}$,
\emph{true} volatility $v_t$, 
and the cliquet's payoff if 
it matured at time $t$.
We process inputs $I_t$ and $u_t$ linearly
via
${W^{I} I_t + W^{u} u_{t - 1} + b}$
with weight matrices $W^{I}, W^{u}$
and bias $b$.
This layer is
followed by four residually stacked blocks,
where each block applies an RMSNorm~\citep{zhang_2019_root}
and an LSTM cell of dimension $32$.
Finally, we pass outputs
to a linear layer with output dimension $d$,
a symexp activation~\citep{hafner_2023_maste},\footnote{
$\text{symexp}(x) := \text{sign}(x) (\exp{\vert x \vert} - 1)$
for $x \in \R$.
}
and a masking layer that sets
components of output $u_t$ to
zero whenever that hedge is not available.
We carefully initialize the last layer
with zero bias and down-scaled (factor of $1\text{e-}3$)
\citep{andrychowicz_2020_what},
He\nobreakdash-initialized
\citep{he_2015_delvi}
weights.

We compare our second-order scheme
to Adam.
For both methods, we grid search hyperparameters
and report the best choices below.
We find that good
architectural and initialization choices (see above) generally
benefit both algorithms,
though DH\nobreakdash-KFAC is more robust to normalizations
\citep[see also][]{ba_2017_distr}.
We increase Adam's learning rate linearly
during the first epoch
and decay it exponentially afterwards
(see values in \cref{fig:loss}).
Similarly, DH-KFAC's trust-region size is
decayed with $\beta^{\text{TR}} = 0.997$ each step
and initialized at
$\trustregion_0 = 1\text{e-}3$.
We train both methods with a batch size of $2048$.
We find that Adam requires relatively
large batch sizes,
which we attribute to the ill-conditioning
of the pathwise loss 
(\cref{subsec:optimizing_hedging_policy}).
KFAC is known to work well with large batch
sizes \citep{ba_2017_distr}
and is therefore well-suited to deep hedging.
For Adam, we clip gradients
such that $\norm{\nabla \loss(\theta)}_2\leq 1$.
For DH-KFAC,
we set
${\varrho = 5\text{e-}4}$,
$\beta^\text{mom} = 0.92$,
and
$\beta^D = \beta^F = 0.95$.

\subsection{Results}\label{subsec:results}

\paragraph{Optimization performance}

\Cref{fig:loss}
visualizes the per-iteration
optimization progress.
We stop training
once a given validation loss target
is reached.
DH\nobreakdash-KFAC progresses
significantly faster
than Adam and only requires
about $25\%$ of Adam's number of steps to reach
the targeted loss level.
We find that for our small network
the per-iteration speedup
decreases total training wall-clock time,
despite our DH\nobreakdash-KFAC implementation
not being optimized.
For larger networks,
distributed computing and performance optimizations
are important
\citep{ba_2017_distr, ueno_2020_rich}.
In deep hedging, the costs of forward- and backward pass
are large relative to the number of parameters,
so we expect the
KFAC overhead to be smaller
than in non-recurrent architectures.\footnote{
For perspective, \citet{ueno_2020_rich} reduce
KFAC's update cost to a multiple of
$1.89$ of SGD's cost
in distributed
ResNet50
training.
}
In our experiments,
we amortize the KFAC computations at rates
$N^{\text{cov}}\,{=}\,5,$ $N^{\text{evd}}\,{=}\,25$.

\begin{figure}
\centering
\includegraphics{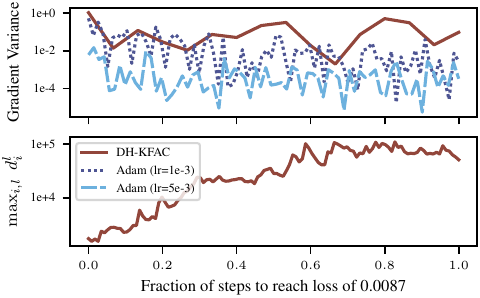}
\Description{
TODO: description 
}
\caption{
Gradient variance and
largest eigenvalue.
}
\label{fig:ekfac_scales}
\end{figure}

The speedup in per-iteration progress suggests
that our second-order scheme can
better deal with the loss function's curvature.
The top plot in \cref{fig:ekfac_scales}
shows the variance of a batch of
\emph{unconditioned} single-sample gradients,
i.e., the trace of their empirical covariance matrix.
We see that Adam's optimization path
goes through regions of the loss surface
with lower gradient variance than DH\nobreakdash-KFAC's.
The bottom plot in \cref{fig:ekfac_scales}
shows that the largest eigenvalue
of our preconditioner
increases for most of
of training.
These observations seem to indicate
that DH\nobreakdash-KFAC allows for
entering high-curvature regions
of the loss surface.
We leave a precise analysis
in terms of \emph{preconditioned} sharpness
\citep{cohen_2024_adapt}
for future work.

\paragraph{Learned hedging policy}

\begin{figure}
\centering
\includegraphics{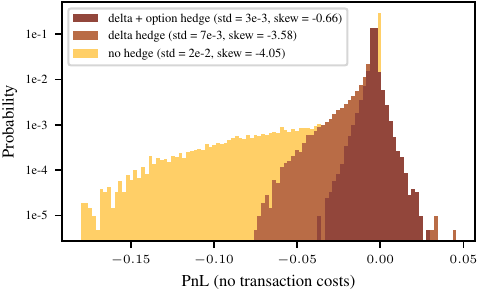}
\Description{
TODO: description 
}
\caption{
PnL histogram (log scale). 
}
\label{fig:histogram}
\end{figure}

\begin{figure}
\centering
\includegraphics{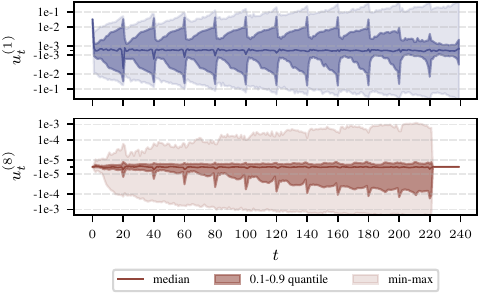}
\Description{
TODO: description 
}
\caption{
Delta hedge (top)
and single
option hedge (bottom).
}
\label{fig:delta_and_opt_hedge}
\end{figure}
We analyze the learned hedging strategy
on an independent test set of $70,000$ paths.
Here, the model was DH\nobreakdash-KFAC-trained,
but we eventually reach the
same conclusions for Adam-trained models.
\Cref{fig:histogram} visualizes the distribution
of the terminal PnL ${(\pf_T - \psi(x))}$
and compares it to 
the unhedged PnL
and 
the PnL obtained when
removing all option hedges.
The learned hedging policy
significantly reduces PnL variance,
primarily with delta hedging.
Including option hedging reduces
the PnL's standard deviation by
another $50\%$
and increases the skewness
of the PnL distribution significantly:
the model hedges in vanilla options to
protect against
tail scenarios.
\Cref{fig:delta_and_opt_hedge}
reinforces this conclusion.
It visualizes the distribution of the
delta- and an option hedge over
time.
The visualized option hedge
is
in a $20$\nobreakdash-step call option with
relative strike of $1.01$,
which lines up closely with the
cliquet's $20$\nobreakdash-step return cap at $0.015$.
We see that hedging activity in both
instruments increases around the reset dates.
For the call, this is when its maturity
aligns with the next cliquet fixing date.

\section{Conclusion}

We proposed a second-order
scheme for optimizing
deep hedging models.
We demonstrated that preconditioning gradients
with a block-diagonal, Kronecker-factored
approximation of the generalized Gauss Newton
matrix significantly speeds up
per-iteration optimization progress.
Beyond the practical use of our algorithm,
our paper provides a new perspective
on standard deep hedging optimization.
This lens of curvature
may for instance be useful
in the design of whitening layers
\citep{desjardins_2015_natur}
as we find our algorithm to still perform well
when setting $G = \eye$.
Beyond deep hedging, our approach
of leveraging second-order information through
a differentiable environment
is, with modification of the preconditioner,
applicable to
other differentiable control and RL problems.

\bibliographystyle{ACM-Reference-Format}
\bibliography{extracted_ref}

\appendix
\vspace{3cm}
\section*{Disclaimer}
This paper was prepared for information purposes by the Quantitative Research group of JPMorgan Chase \& Co and its affiliates (“JP Morgan”), and is not a product of the Research Department of JP Morgan. JP Morgan makes no representation and warranty whatsoever and disclaims all liability, for the completeness, accuracy or reliability of the information contained herein. This document is not intended as investment research or investment advice, or a recommendation, offer or solicitation for the purchase or sale of any security, financial instrument, financial product or service, or to be used in any way for evaluating the merits of participating in any transaction, and shall not constitute a solicitation under any jurisdiction or to any person, if such solicitation under such jurisdiction or to such person would be unlawful.

\end{document}